\documentclass[slac_one]{revtex4}
\usepackage{graphicx}
\usepackage{fancyhdr}

\begin{document}

\title{Determination of Wtb anomalous coupling at the ILC} %% Paper title goes here


\author{E. Devetak}
\affiliation{University of Oxford, UK}

\begin{abstract}
We present a study of experimental determination of the Wtb anomalous couplings in the $t\bar{t}$ process at the International Linear Collider in the framework of effective field theory. The theory predicts several observables that are particularly sensitive to the couplings including the asymmetry of the bottom quark angular distributions. These observables require excellent identification capabilities for the b quark in a detector. We describe tools needed for the identification with particular emphasis on determination of the quark charge and b-tagging together with generic selections for the $t\bar{t}$ events. The focus of the study is on the all hadronic decay channel which provides the largest statistics. In the study we demonstrate advantages of the clean environment of the ILC that allow a precise measurement of the observables. We also explain how this process can be used for optimisation of ILC detector in the context of the SiD detector concept. 

\end{abstract}

\maketitle

\thispagestyle{fancy}

\section{MOTIVATION} % Section title should be in all capitals.
It is known that the top quark is substantially more massive than the rest of the observed quarks. It is therefore worth to search for the physical mechanism that introduces the mass difference and at the same time use the top as a probe into new physics, which often couples to mass. In the presence of new physics it is expected that the coupling of the top would deviate from the standard model predictions. 

In particular one can construct a CP conserving effective Lagrangian of the Wtb vertex and parametrise the anomalous couplings with form factors. It is then possible to identify a set of physical observables \cite{Boos:1999ca}. This analysis focuses specifically on the forward backward asymmetry of the b quarks from all hadronic $t\bar{t}$ decays. To calculate the asymmetry one needs first to identify the b quark jets, discriminate the $b$ and $\bar{b}$ quarks and identify $t\bar{t}$ events.

\section{TAGGING THE BOTTOM QUARK}
After the event reconstruction and the jet finding algorithms, in which all events are artificially forced into a six jet configuration, have been performed the analysis aims to identify all the jets that originated from b quarks. 

\begin{figure}[h]
\begin{minipage}[b]{0.4\linewidth} % A minipage that covers half the page
\centering
\includegraphics[width=7cm, height=4.8cm]{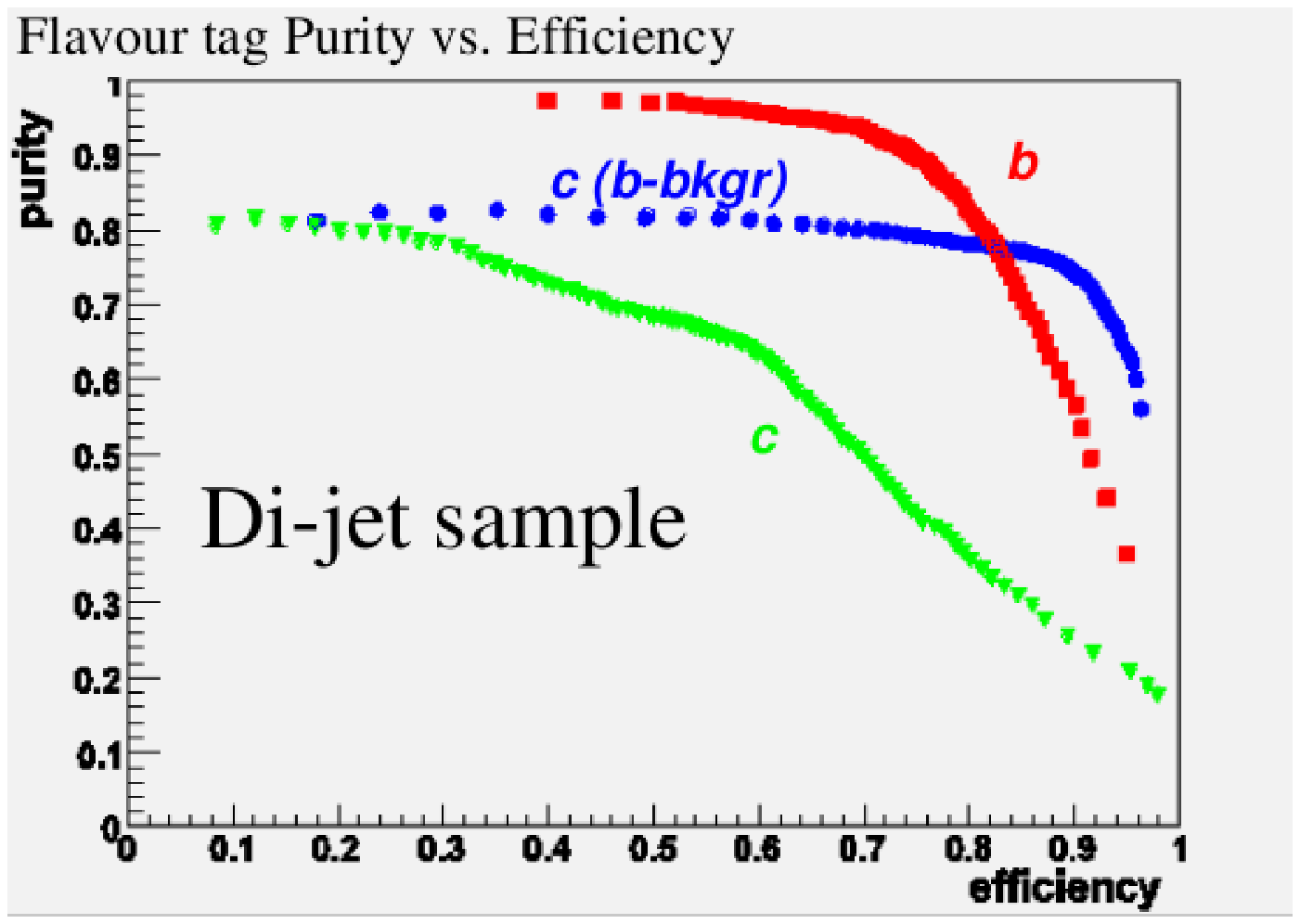}
\caption{Purity vs  Efficiency of flavour tagging in a dijet sample at 500GeV center of mass.}%
\label{fig:btag2}%
\end{minipage}
\hspace{0.3cm} % To get a little bit of space between the figures
\begin{minipage}[b]{0.4\linewidth}
\centering
\includegraphics[width=7cm, height=4.8cm]{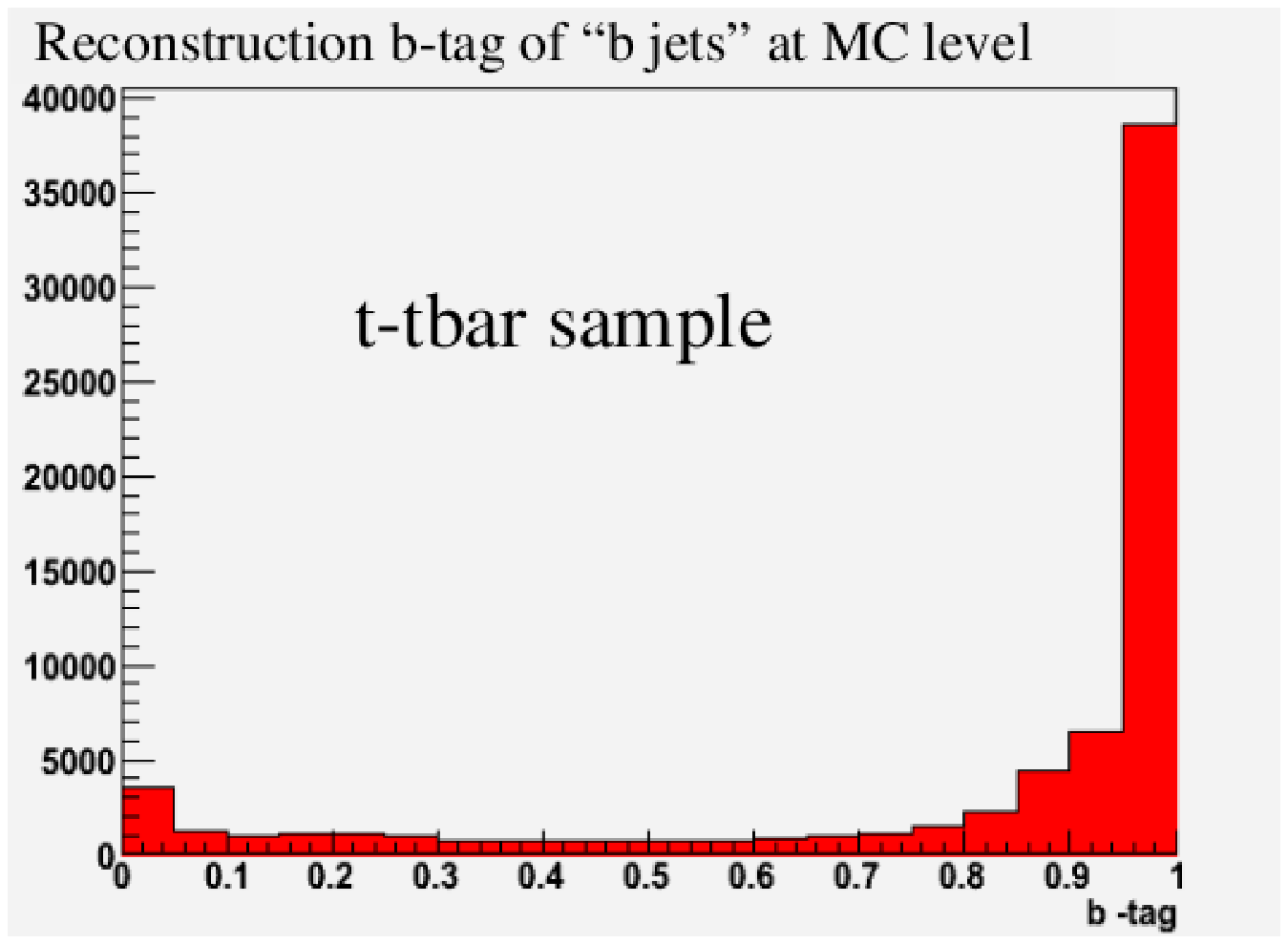}
\caption{Neural Network output for b jets at Monte Carlo level in a $t\bar{t}$ sample.}%
\label{fig:btag6}%
\end{minipage}
\end{figure}

For this task the LCFIVertex Flavour tagging package has been used\cite{LCFI}. The procedure uses the ZVTOP-ZVRES~\cite{Jackson:1996sy} vertex finder followed by a series of algorithms calculating flavour discriminating variables and a neural network, which uses these discriminating variables to determine the flavour of the jet. The results produced from this algorithm in the benchmarking di-jet sample can be seen in  Figure~\ref{fig:btag2}. The results for the $t\bar{t}$ events can be seen in Figure ~\ref{fig:btag6}.

\section{RECONSTRUCTION OF THE QUARK CHARGE}
It is then possible to discriminate between $b$ and $\bar{b}$ quarks. In order to do this two algorithms are employed: the momentum weighted Jet Charge and the momentum weighted Vertex Charge. The algorithms are conceptually the same but they differ in the track selection. In the Vertex Charge algorithm all the tracks present in all the reconstructed secondary vertices are selected; differently in the Jet Charge all the tracks present in the jet are considered. After the track selection is completed we weight the tracks with respect to their momentum since the most energetic tracks carry are more likely to more information about the charge of the $b$ quark. 

\begin{figure}[h]
\begin{minipage}[b]{0.4\linewidth} % A minipage that covers half the page
\centering
\includegraphics[width=7cm, height=4.8cm]{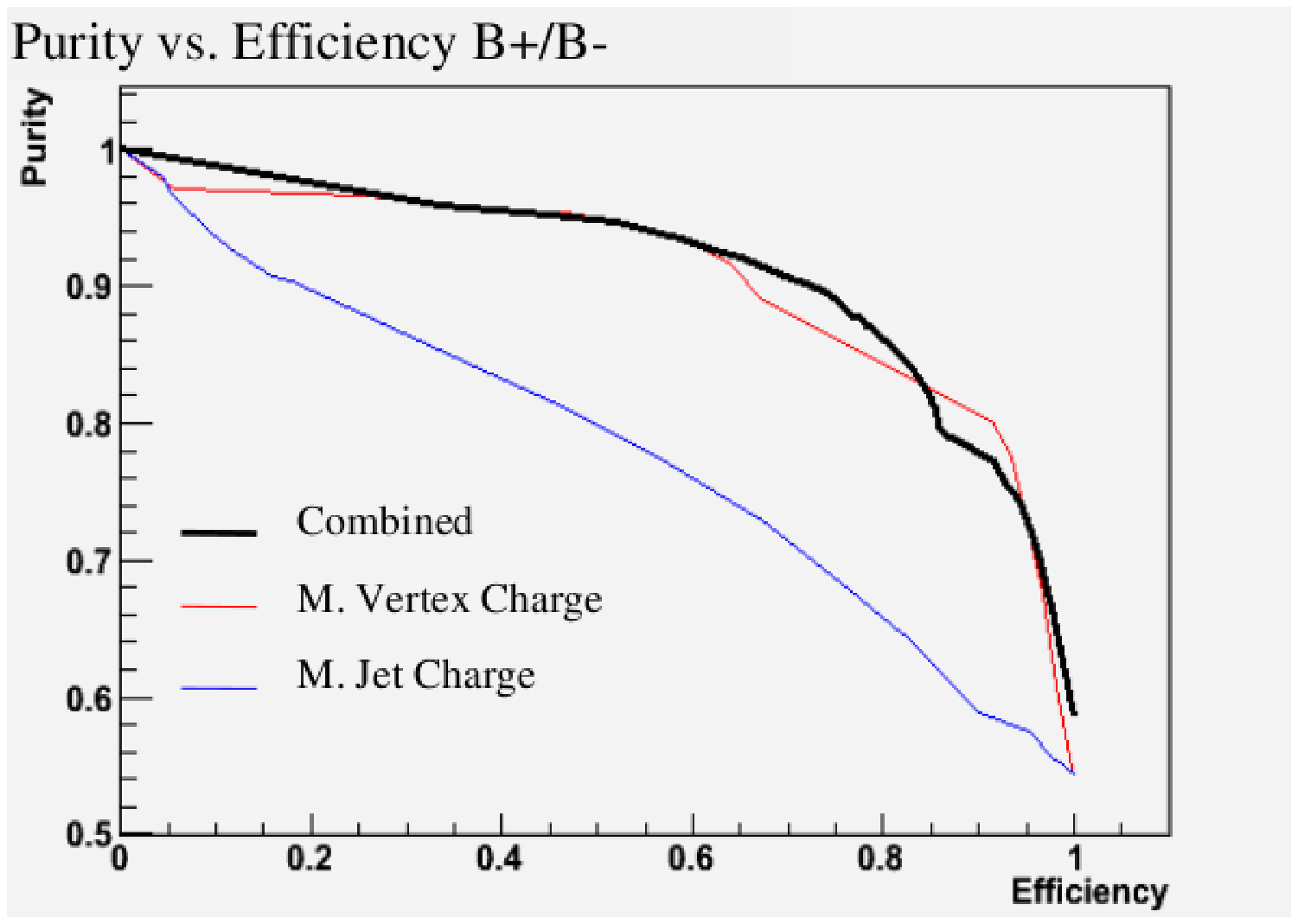}
\caption{Purity vs  Efficiency of the quark charge reconstruction methods in a $B^{+}/B^{-}$ sample.}%
\label{fig:bplus}%
\end{minipage}
\hspace{0.3cm} % To get a little bit of space between the figures
\begin{minipage}[b]{0.4\linewidth}
\centering
\includegraphics[width=5cm, height=4.8cm]{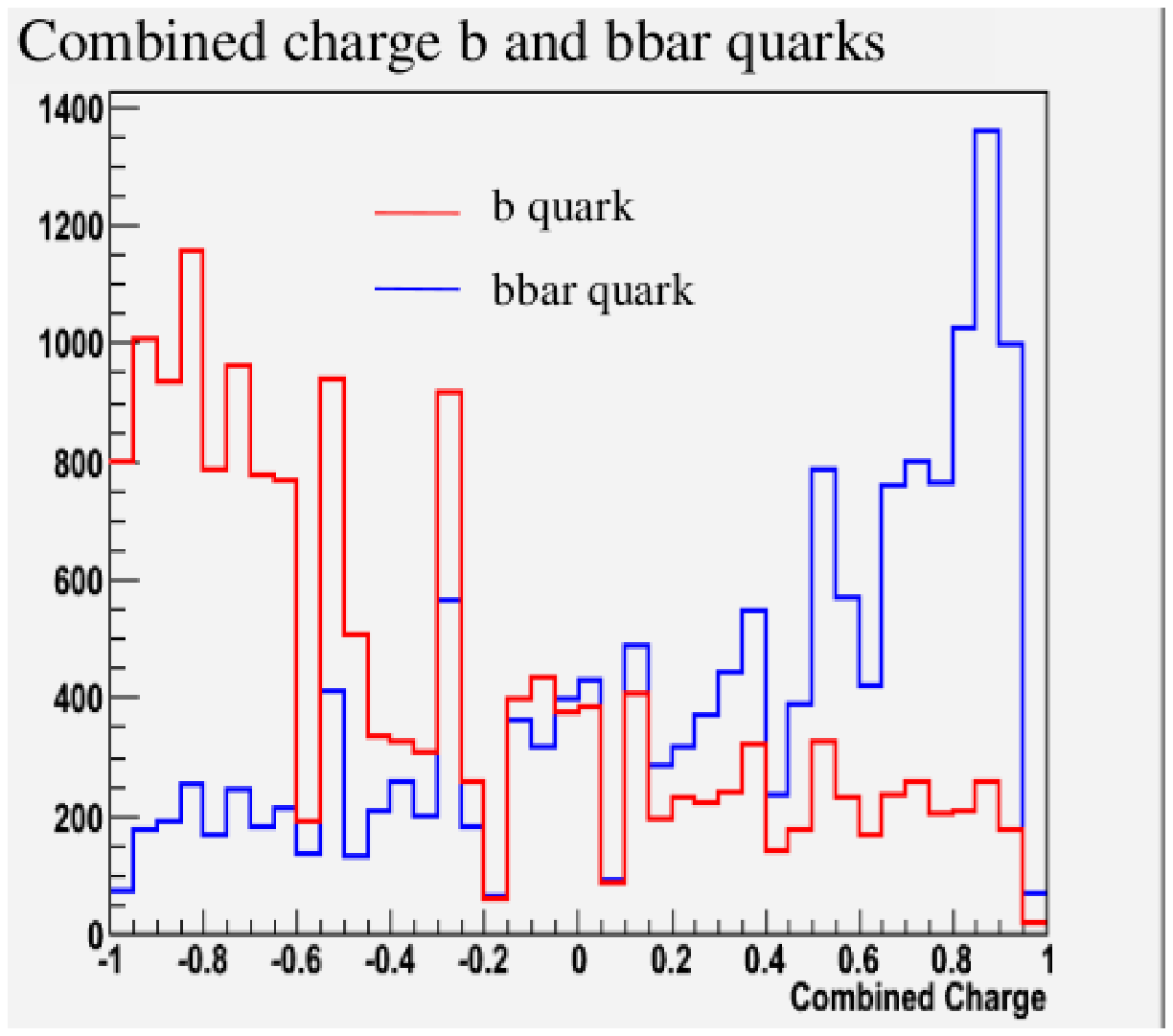}
\caption{Quark charge discriminant from weighted Vertex Charge and Jet Charge.}%
\label{fig:bbar}%
\end{minipage}
\end{figure}

The Vertex Charge method is particularly good at discriminating $B^{+}$ and $B^{-}$ mesons (and hence the constituent quarks) when at least one secondary vertex is reconstructed (See Figure ~\ref{fig:bplus}). This is by far the cleanest of the possibilities. The Jet Charge is instead useful in the more difficult and more frequent occurrences when either the hadron is a $B^{0}$ or $\bar{B^{0}}$ or when a secondary vertex has not been reconstructed. 

After both the Jet Charge and the Vertex charge are calculated it is possible to recombine them into a single discriminating value by using the probability distribution function derived at Monte Carlo level \cite{Abazov:2006qp} (Figure ~\ref{fig:bbar}).

\section{TOP QUARK IDENTIFICATION}

\begin{figure}[h] 
\begin{minipage}[b]{0.4\linewidth} % A minipage that covers half the page
\centering
\includegraphics[width=5cm, height=4.8cm]{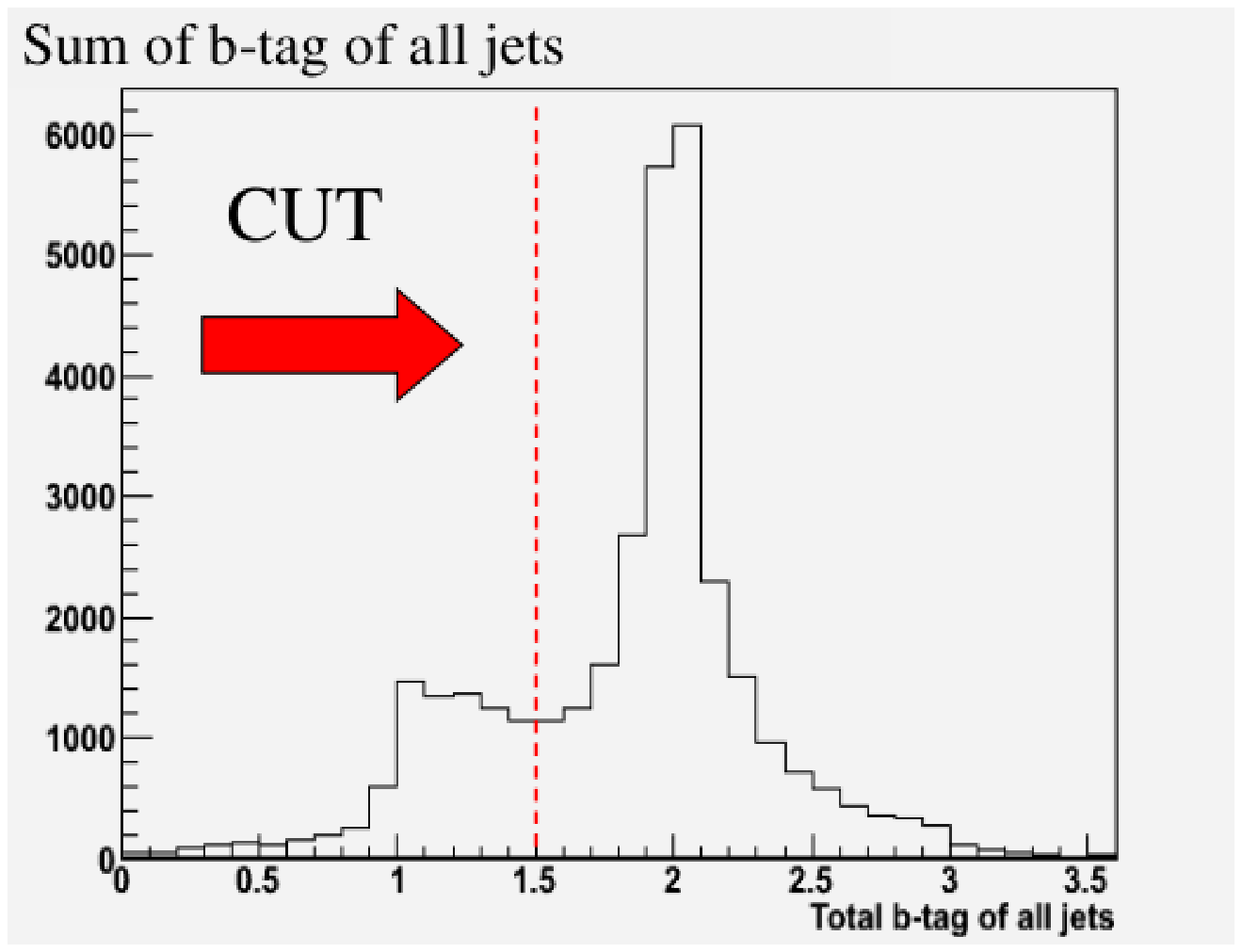} 
\caption{Sum of Neural Network outputs for all jets in the event, with illustrated cut.}%
\label{fig:bcut}%
\end{minipage}
\hspace{0.3cm} % To get a little bit of space between the figures
\begin{minipage}[b]{0.4\linewidth}
\centering
\includegraphics[width=7cm, height=4.8cm]{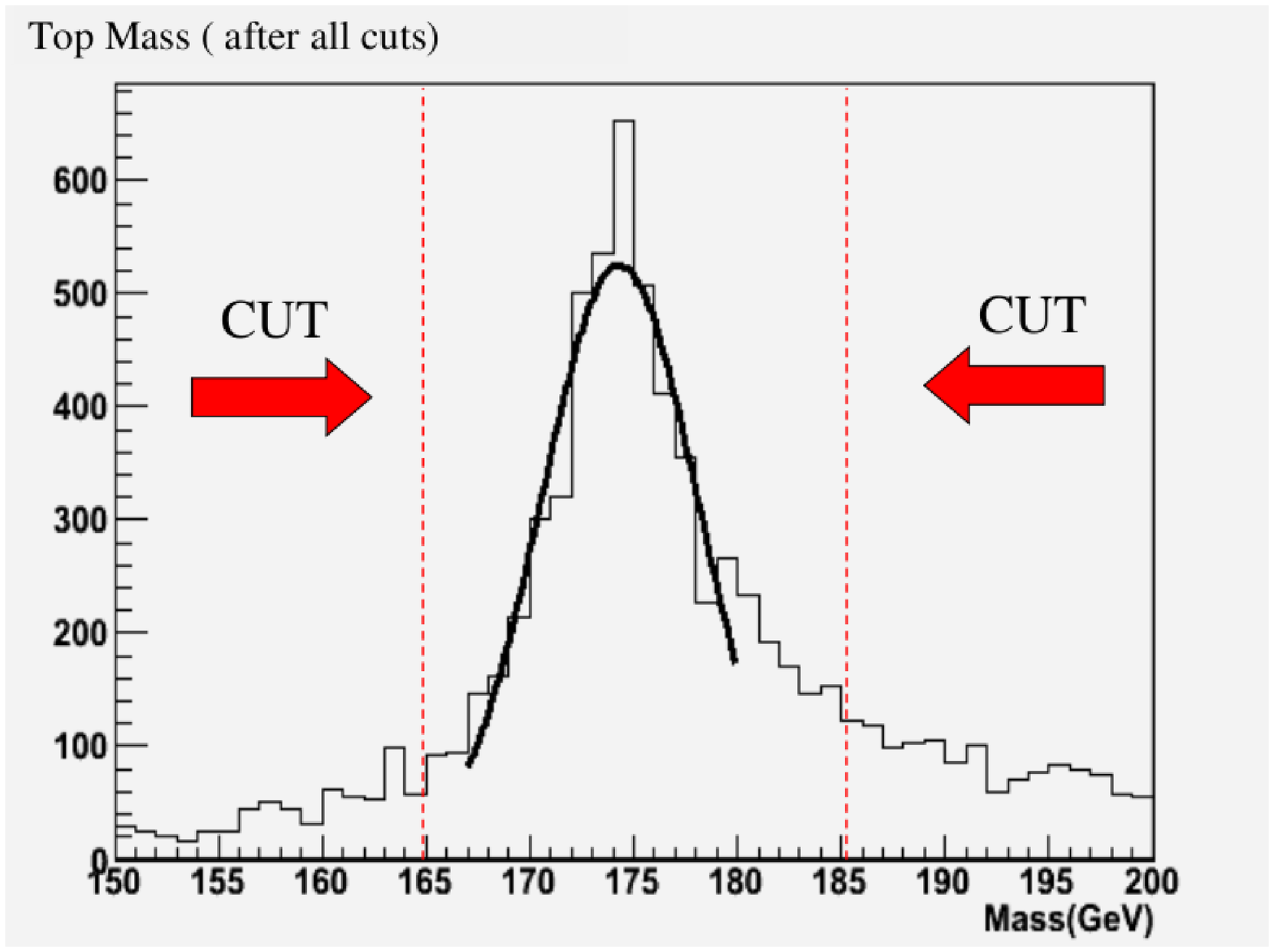}
\caption{Top Mass after reconstruction and kinematic fitting, with illustrated cuts.}%
\label{fig:mass}%
\end{minipage}
\end{figure}

It is at this point essential to look at a method to identify well reconstructed hadronically decayed $t\bar{t}$. This is meant to discriminate both the standard model background, at present not included in the analysis, and the non fully hadronically decaying $t\bar{t}$ events. By applying a set of cuts on number of reconstructed b jets (Figure ~\ref{fig:bcut}.), minimum y-cut of the event jet clustering algorithm, missing energy and mass of reconstructed W as well as other parameters, one is left with what can be considered as well reconstructed hadronic  $t\bar{t}$'s. At present there is no cut on muon and electron identification, however these should be added in the near future.   

On these remaining events a kinematic fit is performed with constraints on total momentum, total energy and masses of the two W's. This produces a  top mass peak (Figure ~\ref{fig:mass}.), which is used to perform the final selection of reconstructed hadronic  $t\bar{t}$ events. 
The efficiency of this reconstruction is approximately 24\% with a purity of 97\%, where the remaining events come from the leptonic decay channels.

\section{BOTTOM QUARK FORWARD BACKWARD ASYMMETRIES}

\begin{figure*}[h]
\centering
\includegraphics[width=70mm]{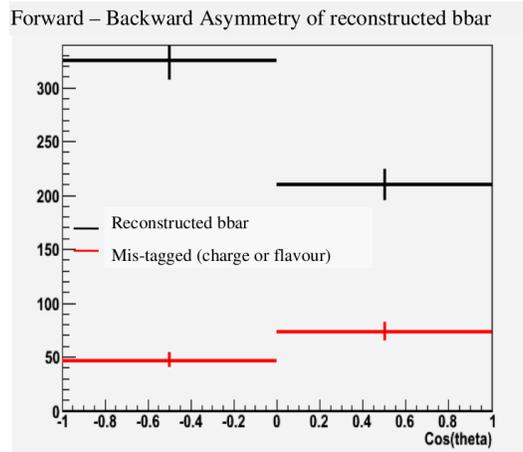}
\caption{Events used in the calculation of Forward Backward Assymetry of $\bar{b}$ quarks, with error deriving from mistagging flavour and charge. For the the precise definition of forward backwards assymetry refer to \cite{Boos:1999ca}.} \label{fig:asymmetry}
\end{figure*}

The events that passed all the top identification criteria and that have a well tagged  $b$ and $\bar{b}$ jets are used to plot the b quark forward backward asymmetry. The calculated asymmetries are  0.33 +/- 0.07 for the $\bar{b}$ (Figure ~\ref{fig:asymmetry}.) quark and 0.14 +/- 0.09 for the $b$ quark. Combining the results one gets 0.26 +/- 0.06; a value consistent with the inputted standard model sample (0.28) \cite{Boos:1999ca}. The results are calculated in the centre of mass frame and hence are a combination of production and decay asymmetries. By calculating the  b quark asymmetry in the top quark frame one can deconvolute the two contributions.

\end{document}